\documentclass{aastex}          
\usepackage{spr-astr-addons}    
\usepackage{epsfig}
\usepackage{graphicx}
\usepackage{color}

\begin{document}
%
\title{The Possibility of Kelvin-Helmholtz Instability in Solar Spicules}

\shorttitle{Kelvin-Helmholtz Instability in Solar Spicules}
\shortauthors{Ajabshirizadeh et al.}
\author{A.~Ajabshirizadeh}
\affil{Research Institute for Astronomy and Astrophysics of Maragha (RIAAM), Maragha 55134-441, Iran}
\and
\author{H.~Ebadi}
\affil{Astrophysics Department, Physics Faculty, University of Tabriz, Tabriz, Iran}
\and
\author{R.~E.~Vekalati}
\affil{Research Institute for Astronomy and Astrophysics of Maragha (RIAAM), Maragha 55134-441, Iran \\
e-mail: \textcolor{blue}{r-vekalati@riaam.ac.ir}}
\and
\author{K.~Molaverdikhani}
\affil{Astrophysics and Planetary Sciences Department, University of Colorado Boulder, Boulder, CO, USA}
\begin{abstract}
Transversal oscillations of spicules axes may be related to the propagation of magnetohydrodynamic waves along them.
These waves may become unstable and the instability can be of the Kelvin-Helmholtz type. We use the dispersion
relation of kink mode derived from linearized magnetohydrodynamic equations. The input parameters of the derived dispersion equation, namely, spicules and their ambient medium densities ratios as well as their corresponding magnetic fields ratios, are considered to be within the range $0-1$. By solving the dispersion equation numerically, we show that for higher densities and lower magnetic fields ratios within the range mentioned, the KHI onset in type $\textsc{ii}$ spicules conditions is possible. This possibility decreases with an increase in Alfv\'{e}n velocity inside spicules. A rough criterion for appearing of Kelvin-Helmholtz instability is obtained. We also drive a more reliable and exact criterion for KHI onset of kink waves.
\end{abstract}
\section{Introduction}
\label{sec:intro}
Coronal heating, the rapid increase in the temperature of the solar corona up to mega kelvins is still a challenging problem of solar physics. Since its discovery in 1943 by Edl\'{e}n \citep{Edlen1943}, various theories have been suggested to explain this phenomenon. One dominant theory attributes coronal heating to the mechanical energy of sub-photospheric motions which is continually transported into the corona by waves and dissipated somehow leading to the energy deposition \citep{Tem2009}. \\
Acoustic waves as the prime candidate for transporting energy to the corona were discarded when it was shown that the energy flux carried by them is too small to have noticeable contribution to the heating of corona and that solar magnetic fields are often concentrated in hot regions.  \citep{Athay1982, Holl1990}. So another scenario, MHD waves, which are supported by magnetic and density characteristics of solar atmosphere were suggested as the energy transporters \citep{Tem2009, Holl1990, Matsumoto2010, Suzuki2006}.\\
 Dynamic chromosphere is dominated by various fine features among which spicules with concentrated magnetic fields and mass densities can be considered as appropriate media for MHD waves and behave as wave guides to carry energy from the photosphere to the corona. MHD waves propagating along spicules may become unstable and the expected instability can be of the Kelvin-Helmholtz type leading to ambient plasma heating \citep{Zhel2012}. The three phases including energy generation by sub-photospheric oscillations, carrying the energy by MHD waves propagating along spicules and energy dissipation by Kelvin-Helmholtz instability are still the subjects of research.\\
Spicules are chromospheric jet-like structures shooting up plasma from the photosphere to the corona. They are observable at the limb of the Sun in chromospheric spectral lines mainly $H_{\alpha}$ and \mbox{Ca\,\textsc{ii}} H-line. Since their discovery in 1877 by Secchi, physical properties and dynamical behaviors  of spicules have been studied extensively \citep{Bec68, Bec72, Ster2000, Tem2009, Tsiro2012, Campos1984}. \citet{De2007} identified another type of spicules with different dynamic properties and called them type $\textsc{ii}$ spicules to be distinguished from classical or type $\textsc{i}$ spicules. Although the general properties of type $\textsc{i}$ spicules vary from one to another, they have diameters of about $700-2500$~km and lengths of about $5000-9000$~km. Their lifetimes lie within 5-15 min, the temperatures are estimated as $9000-17000$~K \citep{Bec68,Bec72,Tem2009}. The temperature of the ambient medium of spicules, chromosphere, is reported in the range of $10^{4}-10^{6}$ K. On the other hand, type $\textsc{ii}$ spicules have shorter lifetimes about $10-150$~s, smaller diameters ($\leq200$~km) and lengths from 1000 km to 7000 km. The short lifetimes of type $\textsc{ii}$ spicules may be either due to their rapid diffusion or heating and their rapid disappearance can be associated with the Kelvin-Helmholtz instability \citep{Tem2010}.\\
Among dynamical behaviors, oscillatory motions as evidences for MHD waves propagating along spicules have been further uncovered through imaging and spectroscopic observations from space and ground based facilities. Observation of kink modes which were accompanied with transverse oscillations of spicules axes was first reported by \citet{Kukh2006}. \citet{Kukh2006, Tem2007} by analyzing the height series of $H\alpha$ spectra in solar limb spicules observed their transverse oscillations with the estimated periods of $20-55$ s and $75-110$ s. \citet{Ebadi2012,Ebadi2013} based on \emph{Hinode}/SOT observations estimated the oscillation period of spicule axis around $90$ and $180$ s. \citet{Tem2007} using doppler shift-oscillations method measured the periods of spicular oscillations that could be due to the kink waves propagation in spicules. Estimates of the energy flux carried by MHD (kink or Alfv\'{e}n) waves in spicules and comparisons
with advanced radiative magnetohydrodynamic simulations indicate that such waves are energetic enough possibly to heat the quiet corona \citep{De2007,Ebadi2012}. The velocities measured for spicules are in the range of about $20-25$~km/s. The new type spicules have higher velocities of order $50-150$~km/s.\\
Hydrodynamic and magnetohydrodynamic flows are often subject to Kelvin-Helmholtz instability (KHI). In hydrodynamics KHI was first described by Kelvin (1871) and Helmholtz (1868). It results from velocity shears between two fluids moving across an interface with a strong enough shear to overcome the restraining surface tension force \citep{Drazin1981}. In magnetohydrodynamics the instability is generated by shear flows in magnetized plasmas. When the flow speed exceeds a critical value, KHI can arise\citep{Chandra1961}. KHI in sheared flow configurations is an efficient mechanism to initiate mixing of fluids, transport of momentum and energy and development of turbulence. Spicules as jet-like structures flowing in ambient plasma environment can be subject to KHI under some conditions. \\
\citet{Ivan2012} studied the conditions under which MHD waves propagating along spicules become unstable because of the Kelvin-Helmholtz instability. He considered spicules as vertical magnetic cylinders containing fully ionized and
compressible cool plasma surrounded in a hot and incompressible plasma environment with density $50-100$ times less than
that of the spicules. Considering mentioned conditions about spicules and their environment he derived the dispersion equation of MHD waves which was sensitive to the parameters $\eta$ (density ratio of outside to that inside of spicule), $b$ (the ratio of the environment magnetic field to that of the spicule) and $M_{A}$ (the ratio of flow speed to Alfv\'{e}n velocity inside spicule) called Alfv\'{e}n-Mach number. With  $\eta=0.01$ and $b=0.36$ and solving the dispersion equation for sausage and kink modes, Zhelyazkov concluded that only kink modes can be affected by the KHI and the critical Alfv\'{e}n-Mach number needed for KHI onset is $M_{Ac}=12.6$ corresponding to jet velocity of order $882$~km/s. For $\eta=0.02$ and $b=0.35$ the critical Alfv\'{e}n-Mach number was $8.9$ meaning that jet speeds of order $712$~km/s or higher are required to ensure that the Kelvin-Helmholtz instability occurs in kink waves propagating along spicules. The conclusion was that these speeds are too high to be registered in spicules. So KHI in cylindrical model of spicules can not happen\citep{Ivan2012}.\\
In this paper we study the possibility of KHI in spicules in a wider range of input parameters $\eta$ and $b$ and apply the actual conditions of spicules. Spicules are considered straight cylindrical jets flowing in ambient plasma environment. The basic of our investigation is the dispersion equations of kink waves derived from linearized MHD equations \citep{Ivan2012,Edwin1983}.

\section{A summary of observational data}
Spicules physical properties as well as their ambient environment characteristics have been widely investigated. Among them densities and magnetic fields play key role in KHI onset through the input parameters $\eta$ and $b$. A summary of observational data on densities and magnetic fields inside and outside spicules are listed in tables \ref{tab1} and \ref{tab2}, respectively.

\begin{table*}[htbp]
\caption{A summary of mass densities and magnetic fields inside spicules}
\label{tab1}
\vspace{0.5cm}
\fontsize{4}{4.8}{\fontsize{4}{4.8}}
\begin{center}
\tabcolsep 1.2pt
\small
\begin{tabular}{|c| c| c| c| c|}
\hline
Height(km) &$n_{i}(cm^{-3})$ &$B_{i}(G)$ &Source &Investigator  \\
\hline
$2000-10000$ &$2.2\times10^{11}-3\times10^{10}$ &$18-76$ &Beckers, $1968, 1972$ &- \\
\hline
$6000$ &$5\times10^{9}-1\times10^{12}$ &- &Beckers, $1972$& \\
\hline
$4000-10000$ &$2.11\times10^{11}-3.5\times10^{10}$ &- &Murawski \& Zaqarashvili, $2013$ &- \\
\hline
- &$10^{11}-10^{12}$ &- & &Krat and Krat, $1971$ \\
$4500$ &$6\times10^{10}$(average) &- & &Alissandrakis, $1973$ \\
$4500$ &$6\times10^{10}-1.2\times10^{11}$ &- &  &Alissandrakis, $1973$   \\
$6000-10000$ &$1.1\times10^{11}-2\times10^{10}$ &- &Tsiropula et al., $2012$ &Krall et al., $1976$ \\
$5000$ &$6\times10^{10}$ &- & &Krall et al., $1976$ \\
- &$2.4\times10^{13}-4.4\times10^{14}$ &$10-76$ & &Kim et al., $2008$ \\
- &- &$10$ \& $30$ & &Lopez Ariste \& Casini, $2005$ \\
- &- &$8-16$ & &Singh and Owivedi, $2007$ \\
- &- &$12-15$ & &Zaqarashvili et al., $2007$ \\
\hline
 &$1.3\times10^{11}$ & & & \\
Base-top &$4.1\times10^{14}-5.5\times10^{13}$ &$20$ &Campos, $1984$ &- \\
 &$1.5\times10^{14}$ & & & \\
\hline
- &$3.3\times10^{14}$ &- &Sterling, $2000$ &- \\
\hline
- &- &$40$ &Centeno et al., $2009$ &Centeno et al., $2009$ \\
\hline
- &- &$10$(quiet sun regions) &Centeno et al., $2009$ &Ramelli et al., $2006$ \\
 & &$50$(active regions) & & \\
\hline
- &- &$18-76$ \& $12-56$ &Yeon-Han Kim et al., $2008$ &Yeon-Han Kim et al., $2008$ \\
\hline
\end{tabular}
\end{center}
\end{table*}

\begin{table*}[htbp]
\caption{A summary of mass densities and magnetic fields outside spicules (chromosphere)}
\label{tab2}
\vspace{0.5cm}
\begin{centering}
\begin{tabular}{|c| c| c| c| c|}
\hline

Region &$n_{e}(cm^{-3})$ &$B_{e}(G)$  &Source &Investigator  \\
\hline
cool corona &$1\times10^{9}$ &$10$  &Aschwanden, $2005$ &- \\
\hline
$2000-10000$~km &$8\times10^{9}-5\times10^{8}$ &-  &Aschwanden, $2005$ &Fontenla et al., $1990$  \\
 & & & &Abriel, $1976$ \\
\hline
- &$10^{9}$ &-  &Zaqarashvili \& Erdelyi, $2009$ &- \\
\hline
quiet sun region &$1\times10^{8}-2\times10^{8}$ &-  & & \\
coronal holes &$5\times10^{7}-1\times10^{8}$ &-  &Aschwanden, $2005$ &-  \\
coronal streamers &$3\times10^{8}-5\times10^{8}$ &-  & & \\
active regions &$2\times10^{8}-2\times10^{9}$ &-  & & \\
\hline
- &$10^{11}-10^{12}$ &-  &Aschwanden et al., $2002$ &- \\
\hline
\end{tabular}
\end{centering}
\end{table*}

It can be seen that despite providing significant data, there are not still definitive and certain results and even sometimes may be conflicting \citep{Tsiro2012, Ster2000}. To study the KHI in spicules conditions and according to the data in tables \ref{tab1} and \ref{tab2} we assume a wide range of values for the input parameters of the dispersion equation, $\eta$ and $b$. This range can be considered between $0$ and $1$.\\
Since $\frac{\rho_{e}}{\rho_{i}}=\frac{n_{e}}{n_{i}}$ we have used number density instead of mass density in the tables.

\section{Theoretical modeling}
\label{sec:theory}
We consider spicule as a vertical cylinder aligned with the z axis of the coordinate system(z is considered perpendicular to the surface of the Sun) with radius $a$ filled with ideal compressible and fully ionized plasma  of density $\rho_{i}$ and pressure $P_{i}$ embedded in a plasma of density $\rho_{e}$ and pressure $P_{e}$. Subindexes i and e refer to internal and external quantities.
Inside (outside) the spicule is immersed in a uniform vertical magnetic field $B_{i}$ ($B_{e}$) directed along z axis.
The plasma inside (outside) the spicule flows along the spicule axis with speed $U_{i}~(U_{e})$. For convenience the ambient medium can be considered as a reference frame. In this case $U=U_{i}-U_{e}$ is the relative flow velocity. The equilibrium condition requires that the total pressure including gas and magnetic pressures inside and outside the spicule balance \citep{Edwin1983}:
\begin{equation}
\label{eq:presure} \ p_{i}+B_{i}^{2}/2\mu=p_{e}+B_{e}^{2}/2\mu ,
\end{equation}
where $\mu$ is the magnetic permeability. The density ratio is obtained as follows:
\begin{equation}
\label{eq:dens} \frac{\rho_{e}}{\rho_{i}}= \frac{2c_{si}^{2}+\gamma v_{Ai}^{2}}{2c_{se}^{2}+\gamma v_{Ae}^{2}},
\end{equation}
where $c_{si}=\sqrt{\gamma p_{i}/\rho_{i}}$ ($c_{se}=\sqrt{\gamma p_{e}/\rho_{e}}$) and $v_{Ai}=B_{i}/\sqrt{\mu \rho_{i}}$ ($v_{Ae}=B_{e}/\sqrt{\mu \rho_{e}}$) are the sound and Alfv\'{e}n speeds inside (outside) the cylinder, respectively and $\gamma$ is the ratio of the specific heats.
We consider the equilibrium situation and perturb the system linearly, then:
$\rho=\rho_{0}+\rho_{1}$, $p=p_{0}+p_{1}$, $\mathbf{v}=\mathbf{U}+\mathbf{v_{1}}$ and $\mathbf{B}=\mathbf{B_{0}}+\mathbf{B_{1}}$,
in which the indices $0$ and $1$ refer to the equilibrium and small perturbed situations, respectively.
 Using the basic equations of compressible, adiabatic, inviscid and ideal MHD including the continuity of mass, momentum, energy, induction and solenoid equations and linearizing them, the ideal linearized MHD equations are as follows:
\begin{equation}
\label{eq:cont} \frac{\partial\rho_{1}}{\partial t}+(\mathbf{U}\cdot \nabla)\rho_{1}+\rho_{0}\nabla\cdot \mathbf{v}_{1}=0,
\end{equation}
\begin{eqnarray}
\label{eq:momentum} \rho_{0}\frac{\partial\mathbf{v_{1}}}{\partial t}+ \rho_{0}(\mathbf{U}\cdot\nabla)\mathbf{v_{1}}+\nabla(p_{1}+ \frac{1}{\mu_{0}}\mathbf{B_{0}}\cdot \mathbf{B_{1}}) \\
\nonumber - \frac{1}{\mu_{0}}(\mathbf{B_{0}}\cdot \nabla )\mathbf{B_{1}}=0,
\end{eqnarray}
\begin{equation}
\label{eq:energy} \frac{\partial p_{1}}{\partial t}+ (\mathbf{U}\cdot\nabla)p_{1}+ \gamma p_{0} \nabla\cdot\mathbf{v_{1}}=0,
\end{equation}
\begin{equation}
\label{eq:induction} \frac{\partial \mathbf{B_{1}}}{\partial t}+(\mathbf{U}\cdot\nabla)\mathbf{B_{1}}-(\mathbf{B_{0}}\cdot\nabla)\mathbf{\mathbf{v_{1}}}+\mathbf{B_{0}}\nabla\cdot\mathbf{\mathbf{v_{1}}}=0,
\end{equation}
\begin{equation}
\label{eq:solnoid} \nabla\cdot\mathbf{B_{1}}=0.
\end{equation}
Because the mass density of spicules in their length limits does not change appreciably, the gravity force term has been omitted in equation \ref{eq:momentum}. The effects of gravity stratification, twist and magnetic expansion may be investigated in future works.
By applying boundary conditions according to which the perturbed interface has to be continuous and total pressure including gas and magnetic pressures inside and outside the spicule must balance at the interface, and after some manipulations the dispersion equation of MHD waves for various modes propagating along the spicule is obtained as \citep{Ivan2012, Terra2003, Nakariakov2007}\\
\begin{eqnarray}
\label{eq:totdisper}
\frac{\rho_{e}}{\rho_{i}}(\omega^{2}-k^{2}_{z}v^{2}_{Ae})\kappa_{i} \frac{I'_{m}(\kappa_{i}a)}{I_{m}(\kappa_{i}a)}\\
-[(\omega-\mathbf{k}\cdot\mathbf{U})^{_{2}}-k^{2}_{z}v^{2}_{Ai}]\kappa_{e}
\nonumber \frac{K'_{m}(\kappa_{e}a)}{K_{m}(\kappa_{e}a)}=0,
\end{eqnarray}
where \\
\begin{equation}
\label{eq:kapa} \ \ \kappa^{2}=-\frac{[(\omega-\mathbf{k}\cdot\mathbf{U})^{2}-k^{2}_{z}c^{2}_{s}][(\omega-\mathbf{k}\cdot\mathbf{U})^{2}-k^{2}_{z}v^{2}_{A}]}
{(c^{2}_{s}+v^{2}_{A})[(\omega-\mathbf{k}\cdot\mathbf{U})^{2}-k^{2}_{z}c^{2}_{T}]}
\end{equation}
and
\\
\begin{equation}
\label{eq:tube} c_{T}=\frac{c_{s}v_{A}}{(c^{2}_{s}+v^{2}_{A})^{1/2}}.
\end{equation}
 $I_{m}$ and $K_{m}$ are modified bessel functions of the first and second kind, respectively and $I'_{m}$ and $K'_{m}$ are their derivatives with respect to radial component of cylindrical coordinates.
m stands for various MHD modes and $m=1$ represents the kink mode in the dispersion equation.
$\kappa$ and $c_{T}$ are called wave attenuation coefficient and tube velocity, respectively \citep{Edwin1983}.
Normalizing all velocities in dispersion equation \ref{eq:totdisper} to $v_{Ai}$ yields more appropriate form of the dispersion equation. \\
\begin{eqnarray}
\label{eq:finaldisper}(\eta V^{2}_{ph} -b^{2})\kappa_{i}a \frac{I'_{m}(\kappa_{i}a)}{I_{m}(\kappa_{i}a)}-[(V_{ph}-M_{A})^{2}-1]\kappa_{e}a \\
\nonumber \frac{K'_{m}(\kappa_{e}a)}{K_{m}(\kappa_{e}a)}=0
\end{eqnarray}
in which
$V_{ph}=v_{ph}/v_{Ai}=\omega/(k_{z}v_{Ai})$ is dimensionless phase velocity,
$K=k_{z}a$ is wave number normalized to spicule radius,
$M_{A}=U/v_{Ai}$ is Alfv\'{e}n-Mach number,
$\eta=\rho_{e}/\rho_{i}$ is the density contrast
and $b =B_{e}/B_{i}$ is the ratio of the exterior magnetic field to the interior.
Considering spicule as a cool plasma jet and the environment as a hot and incompressible plasma leads consequently to $c_{s}\rightarrow 0$ and $c_{s}\rightarrow \infty$, respectively and correspondingly the internal and external wave attenuation coefficients change into   $\kappa_{i}a=[1-(V_{ph}-M_{A})^{2}]^{1/2}K$ and $\kappa_{e}a=K$. Using these approximations the dispersion equation \ref{eq:finaldisper} can be more simplified and for kink mode ($m=1$) it takes the form \\
\begin{eqnarray}
\label{eq:kinkdisper}(\eta V^{2}_{ph}-b^{2})[1-(V_{ph}-M_{A})^{2}]^{1/2} \frac{I'_{1}(\kappa_{i}a)}{I_{1}(\kappa_{i}a)} \\ \nonumber -[(V_{ph}-M_{A})^{2}-1] \frac{K'_{1}(K)}{K_{1}(K)}=0 .
\end{eqnarray}

By solving equation \ref{eq:finaldisper} we can obtain phase velocity as a function of wave number $V_{ph}(K)$ in which the real part $Re(V_{ph})$ represents dispersion curve of kink waves and the imaginary part $Im(V_{ph})$ represents the growth rate of unstable kink waves propagating along spicules.

\section{Numerical results and discussion}
\begin{figure}
\centering
\includegraphics[width=8cm]{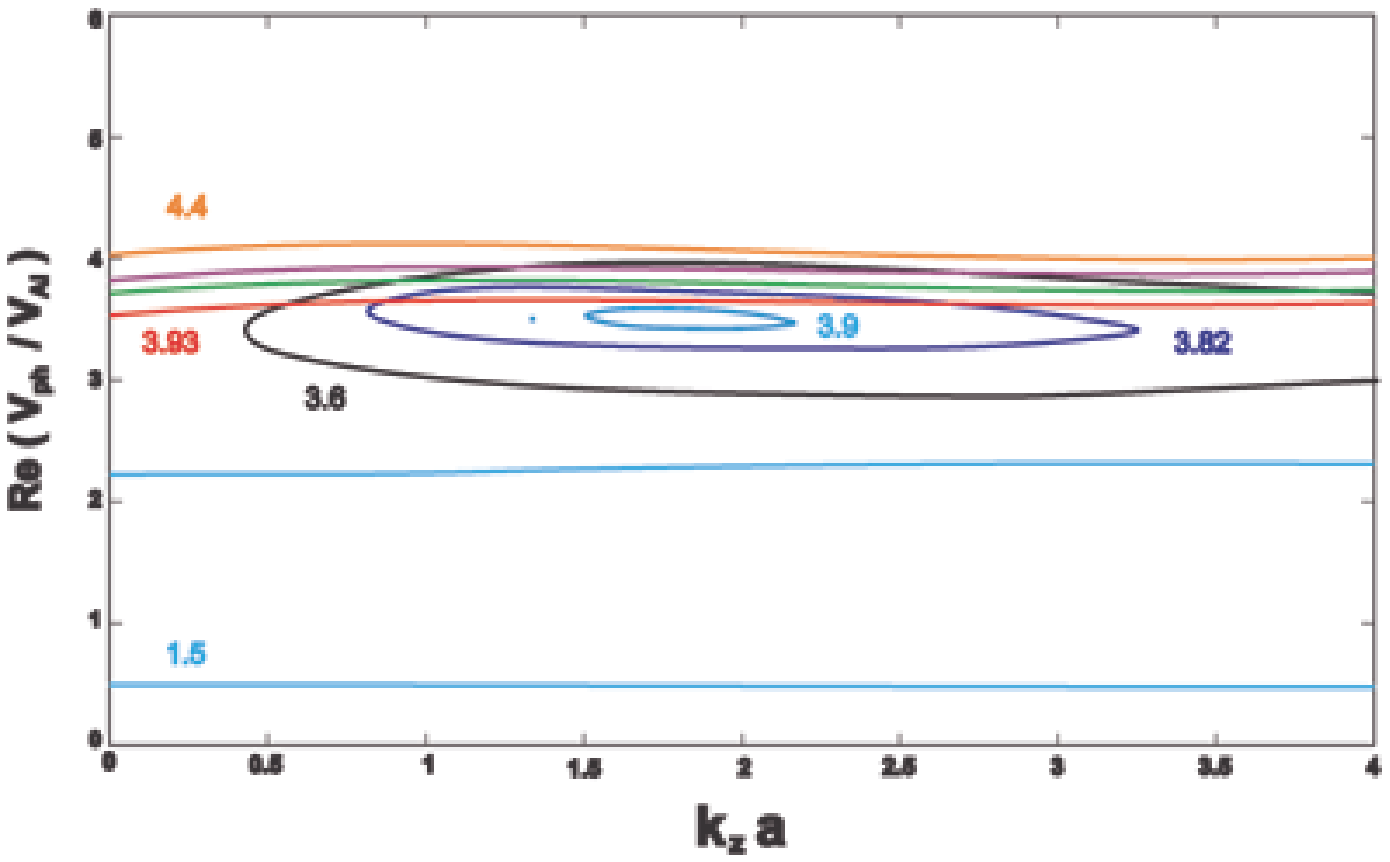}
\includegraphics[width=8cm]{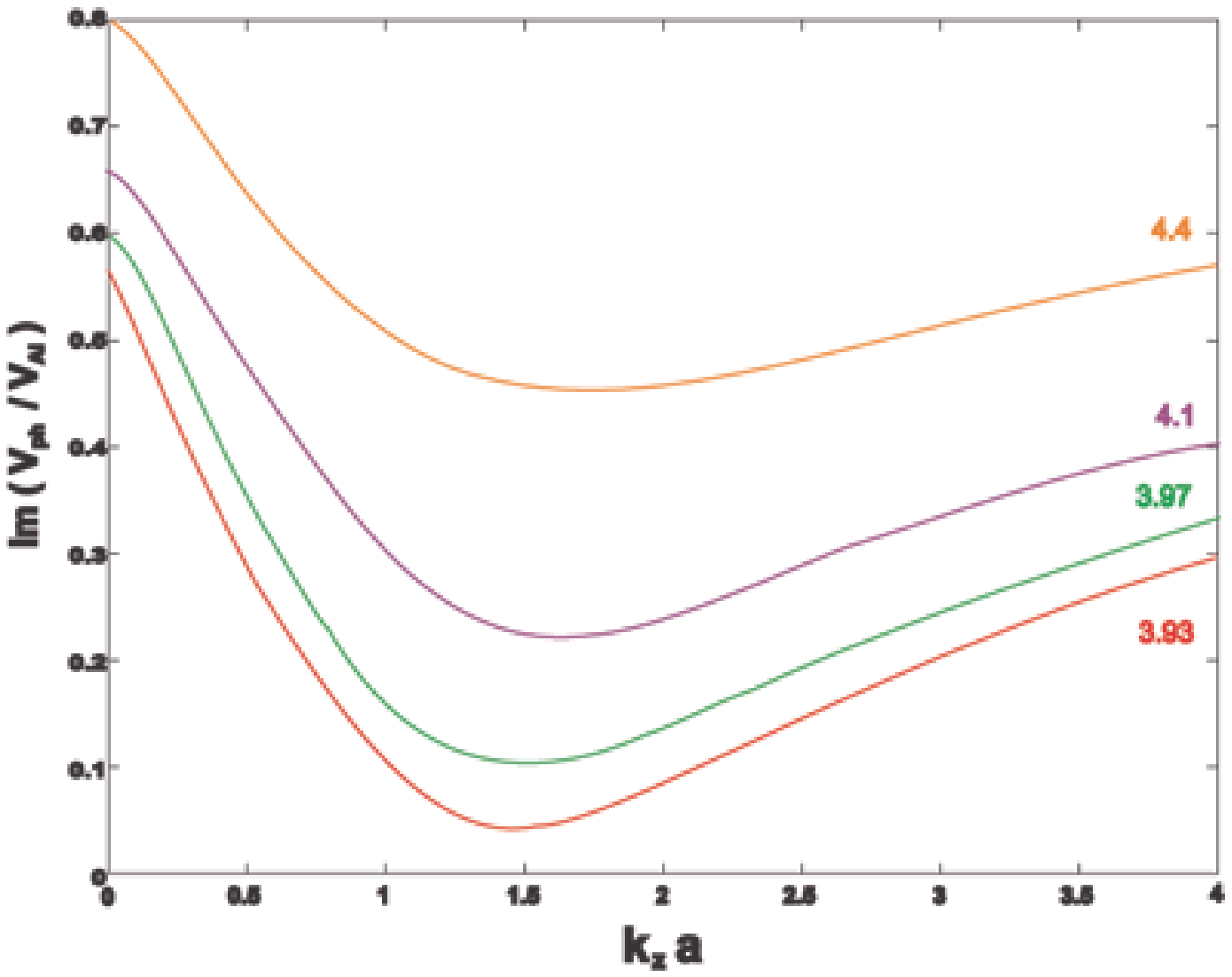}
\caption{Dispersion curves of kink waves propagating along spicules (upper panel) and growth rates curves of unstable kink waves (lower panel) at the input parameters $b=0.25$ and $\eta=0.1$. \label{fig1}}
\end{figure}

First we solved equation \ref{eq:finaldisper} using M\"{u}ller method \citep{muller1956} at various values of the input parameters $\eta$ and $b$. Based on the observational results, they both are considered to be within the range $0-1$. In each case varying Alfv\'{e}n-Mach number $M_{A}$ from zero to some reasonable numbers, yields instability threshold value called critical Alfv\'{e}n-Mach number $M_{Ac}$. When $M_{A}$ reaches the critical value $M_{Ac}$, the system becomes unstable and the instability is of Kelvin-Helmholtz type. The beginning of KHI is accompanied by appearing growth rates curves in the imaginary part of $V_{ph}(K)$ and changing the closed dispersion curves into nearly smooth ones. For input parameters $b=0.25$ and $\eta=0.1$ we obtained $M_{Ac}=3.93$. The dispersion (upper panel) and instability growth rates (lower panel) curves for these parameters are depicted in Figure~\ref{fig1}.

\begin{figure}
\centering
\includegraphics[width=8cm]{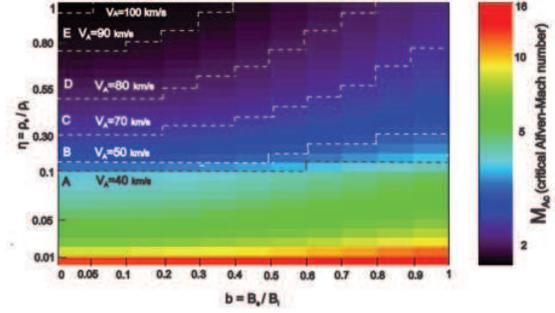}
\caption{Minimum Alfv\'{e}n-Mach numbers in which spicules as cylindrical plasma jets can become unstable of Kelvin-Helmholtz type based on various values of $\eta$ and $b$. The colors identify $M_{Ac}$'s. The determined regions correspond to the range of type $\textsc{ii}$ spicules speeds in various Alfv\'{e}n velocities. \label{fig2}}
\end{figure}

The results of solving equation \ref{eq:finaldisper} at various values of $\eta$ and $b$ within the range $0-1$ are shown in Figure \ref{fig2}. The values of $M_{Ac}$'s are demonstrated by colors. We note that smaller values of $M_{Ac}$'s are concentrated in the regions with lower density contrasts and smaller magnetic fields ratios. These regions are specified by dark blue color. On the other hand red-colored areas present larger values of $M_{Ac}$. These areas correspond to the smaller $\eta$'s and larger $b$'s.

\begin{figure}
\centering
\includegraphics[width=8cm]{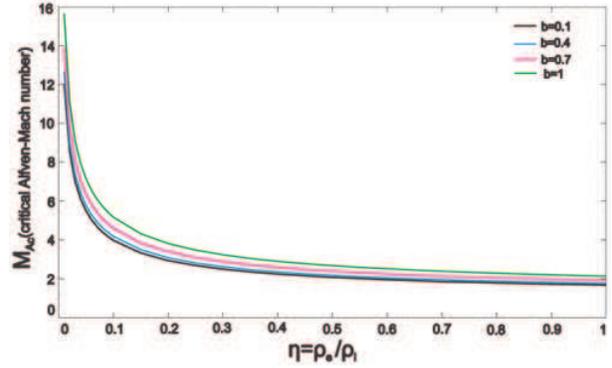}
\caption{The behavior of Alfv\'{e}n-Mach number in terms of the input parameter $\eta$. \label{fig3}}
\end{figure}

\begin{figure}
\centering
\includegraphics[width=8cm]{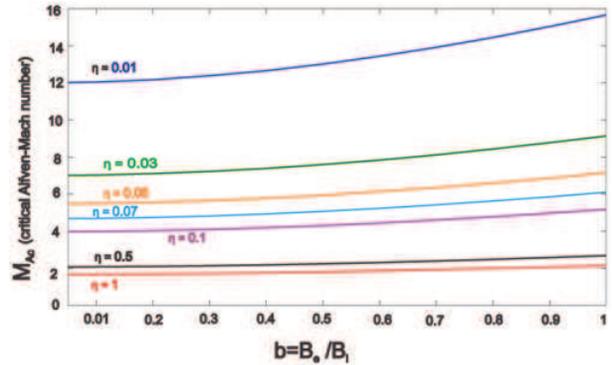}
\caption{The behavior of Alfv\'{e}n-Mach number in terms of the input parameter $b$. \label{fig4}}
\end{figure}

Figure~\ref{fig3} and Figure~\ref{fig4} show the behavior of $M_{Ac}$ with respect to the parameters $\eta$ and $b$, respectively.
According to these figures it can be clearly emphasized that $M_{Ac}$ becomes smaller while $\eta$ increases and $b$ decreases. To discuss the subject more physically, we employ the relation $M_{A}=\frac{U}{V_{Ai}}$. It is obvious that when the system is in critical situation, this relation can be written in the form $U_{c}=V_{Ai} M_{Ac}$ where $U_{c}$ is critical flow speed, the minimum speed at which the flow  becomes unstable. It can be concluded that at smaller values of $M_{Ac}$, lower critical flow speeds are needed for KHI onset. Since the range of type \textsc{ii} spicules velocities are estimated to be between $50$~km/s and $150$~km/s, we are interested in the regions of $b-\eta$ plane (we call them KHI regions) where $M_{Ac}$'s take the values that their corresponding $U_{c}$'s can overlap spicules speeds. On the other hand we know from the aforementioned relation that $U_{c}$ is proportional to $V_{Ai}$ too, therefore to determine critical flow speed of a spicule at a given point $(b,\eta)$, it is necessary to specify Alfve\'{e}n velocity $V_{Ai}$ inside it. Mathematically $V_{Ai}$ is a function of density and magnetic field, therefore one can find a lot of choices for it at the point $(b,\eta)$. To avoid the complexity due to the multiple choices for $V_{Ai}$, we prefer to continue our discussion using various values of Alfv\'{e}n velocities. Considering $V_{Ai}=40$~km/s, the minimum and maximum values of critical flow speed can be obtained as $U_{c min}=60$~km/s and $U_{c max}=625$~km/s corresponding to $\eta=1$, $b=0.05$ and $\eta=0.01$, $b=1$, respectively. The comparison between spicules observed velocities, $(50-150)$~km/s, and their critical flow speeds ,$(60-625)$~km/s shows a partly overlapping. The overlapping region (KHI region) for $V_{Ai}=40 km/s$ is the whole area above dashed line $A$ in Figure \ref{fig2}. For the parameters $b$ and $\eta$ which their values are out of KHI region, the critical flow speeds are higher than those ever observed in spicules and so can be discarded. As a result we can deduce that spicules which their input parameters ($b$,$\eta$) are inside the KHI region and flowing with the minimum speed ($U_{c}$) assigned to that point according to the considered Alfv\'{e}n velocity, can be subject to Kelvin-Helmholtz instability.
By increasing $V_{Ai}$ overlapping between spicules observed and critical speeds gradually decreases and the KHI region becomes smaller indicating that the possibility of occurring KHI diminishes. In Figure \ref{fig2} KHI regions for Alfv\'{e}n velocities of $50$~km/s, $70$~km/s, $80$~km/s, $90$~km/s and $100$~km/s are indicated with areas above dashed lines B, C, D, E and F.
At $V_{Ai}=101$~km/s and higher the minimum critical flow speed exceeds the maximum type \textsc{ii} spicules velocity ($150$~km/s) leading to zero overlap. In other words for $V_{Ai} > 100$~km/s the possibility of KHI in spicules vanishes.\\
A rough criterion for the occurrence of KH instability of kink waves was suggested by Andries \& Goossens \citep{Andries2001}. In our notation it is in the form of
\begin{equation}
\label{eq:Andries}M_{Ac > 1+b/\sqrt{\eta}} .
\end{equation}
We derived another rough criterion for KHI onset of kink waves. Since Bessel functions and their derivatives in the general and simplified equations \ref{eq:finaldisper} and \ref{eq:kinkdisper}, respectively are appeared as the ratios, we realized that these ratios are real valued and moreover
\begin{equation}
\label{eq:Bessel}\frac{I'_{1}(\kappa_{i}a)}{I_{1}(\kappa_{i}a)}=-\frac{\kappa_{e}}{\kappa_{i}}\frac{K'_{1}(\kappa_{e}a)}{K_{1}(\kappa_{e}a)} .
\end{equation}
By these assumptions the resulted criterion is:
\begin{equation}
\label{eq:ours}M_{Ac} > \sqrt{\frac{\eta+1}{\eta}(b^{2}+1)}
\end{equation}
which seems to give more reliable and exact predictions.

\section{Conclusion}
\label{sec:concl}
We considered spicules as cylindrical jets shooting up plasma from the Sun's surface along the $\widehat{z}$ axis of the cylinder. Both spicules and their environment are embedded in magnetic fields directed in $\widehat{z}$ axis. Using linearized MHD equations we obtained the dispersion equation of kink waves propagating along the spicules. The key parameters of the dispersion equation which affect Kelvin-Helmholtz instability occurrence, are $\eta$, $b$ and $M_{A}$. They are defined as $\eta=\rho_{e}/\rho_{i}$, $b=B_{e}/B_{i}$ and $M_{A}=U/V_{Ai}$, where $\eta$ and $b$ are mass densities and magnetic fields ratios of outside to inside spicules and $M_{A}$ is the ratio of spicule streaming speed to Alfv\'{e}n velocity inside it. Observational values of physical properties $\rho_{e}$, $\rho_{i}$, $B_{e}$ and $B_{i}$ reported in the literature are employed to determine the key parameters $\eta$ and $b$. It was shown that they can range from $0$ to $1$. We solved the dispersion equation at various values of $\eta$ and $b$ within the range mentioned, and obtained a critical Alfv\'{e}n-Mach number $(M_{Ac})$ for each ($b,\eta$) pair. The final result is an array of $M_{Ac}$'s representing the threshold values of $M_{A}$'s in which spicules can be subject to KHI. According to the relation $U_{c}=V_{Ai}~M_{Ac}$ and substituting $V_{Ai}=40~km/s$ we can determine a region in the $b-\eta$ plane where the critical flow speeds overlap with the type $\textsc{ii}$ spicules speeds. Out of this region $U_{c}$'s are higher than the observed speeds of spicules and then they can be discarded. Considering larger values for $V_{Ai}$ makes KHI region smaller indicating that the possibility of KHI onset in type $\textsc{ii}$ spicules conditions becomes smaller. For Alfv\'{e}n velocities of $89$~km/s and lower the onset of Kelvin-Helmholtz instability in type \textsc{ii} spicules conditions is possible. This possibility depends on the area of the KHI region. The curves of $M_{Ac}$ in terms of $\eta$ and $b$ show that the threshold value of Alfv\'{e}n-Mach number becomes smaller while increasing $\eta$ and decreasing $b$.\\
Solving the equation \ref{eq:finaldisper} in which spicules and their surrounding environment are considered in more realistic conditions and the exerted approximations are removed, led to critical Alfv\'{e}n-Mach numbers smaller than those resulted from equation \ref{eq:kinkdisper} showing that appearing of KHI gets more possible. \\
To provide more reliable predictions for KHI onset of kink waves, we derived a criterion using Bessel functions properties. It takes the form\\

$M_{Ac} > \sqrt{\frac{\eta+1}{\eta}(b^{2}+1)}$.

\acknowledgments
This work has been supported by the Research Institute for Astronomy and
Astrophysics of Maragha (RIAAM), Maragha, Iran.
\makeatletter
\let\clear@thebibliography@page=\relax
\makeatother

\end{document}